# Self-Consistent Effective Hamiltonian Theory for Fermionic Many Body Systems

Xindong Wang[*] and Hai-Ping Cheng

Quantum Theory Project, Department of Physics, University of Florida

## Abstract

Using a separable many-body variational wavefunction, we formulate a self-consistent effective Hamiltonian theory for fermionic many-body system. The theory is applied to the two-dimensional Hubbard model as an example to demonstrate its capability and computational effectiveness. Most remarkably for the Hubbard model in 2-d, a highly unconventional quadruple-fermion non-Cooper-pair order parameter is discovered.

## Introduction

The holy grail in theoretical condensed matter many-body physics is to overcome the limitations imposed on the Hartree-Fock and/or Kohn-Sham (Kohn & Sham, 1965) type of non-interacting mean-field variational wavefunction (single Slater wavefunction) approach, which cannot adequately address strongly correlated nature of many-fermion systems. Many ideas for going beyond the non-interacting fermions have been proposed with various degrees of success. Most notably are the resonant valence bond picture by Phil Anderson (Anderson, 1987) stemming from the insights of chemical bonds dominating the physical properties, the Green function based dynamic mean field theory (G.Kotliar & Vollhardt, 2004) (Kotliar, Savrasov, Pallson, & Biroli, 2001), and the related Gutzwiller-density-functional theory (Ho, Schmalian, & Wang, 2008), which tries to improve Kohn-Sham theory with some single site correlation taken into account. In parallel, various advanced quantum chemistry methods have been developed to treat strong correlation in molecular systems. However, applications to extended systems are very limited. In this paper, using a separable form of many body wavefunctions between a local system *S* and its environment *E*, we propose a self-consistent theory that fully captures the physics of the local bonds and the self-consistent coherence-preserving mean-field effect from the environment in which the local system is embedded. Our formulation is rigorous and can be applied to realistic systems such as transition metal solids and molecules including bio-molecules that contain reaction centers. In a review article, Chan and his group discussed three possible embedding schemes (Sun & Chan, 2016 ). The last of the three methods discussed by Sun and Chan, the Density Matrix Embedding Theory (DMET) of Knizia and Chan (Knizia & Chan, 2012) (Knizia & Chan, 2013) contains very similar idea in terms of using a separable form of variational wavefunction for the total system. A key difference between DMET and this paper, in our opinion, lies in the effective Hamiltonian for the fragment subsystem in our theory that contains a particle number non-conserving term due to the coherent entanglement of the fragment and the environment it is embedded in. Thus the self-consistency condition for the variational solution is specified differently in DMET compared to our theory.

We will first derive the basic equations of self-consistent effective Hamiltonian theory at zero temperature. The theory is then applied to 2-d Hubbard model, where some exact analytical properties, including superconductivity and anti-ferromagnetism will be discussed. Perhaps most importantly, a highly unconventional quadruple-fermion (non-Cooper pair) order parameter is discovered for 2-d Hubbard model. The finite temperature version of the theory is then given in the end.

## Many-body Ground State at Zero Temperature

We are seeking a variational solution with the following separable wavefunction

$$|\Psi\rangle = \left\{\sum_i c_i |\psi_i^S\rangle\right\}\left\{\sum_\mu c_\mu |\psi_\mu^E\rangle\right\} = |\Psi_S\rangle|\Psi_E\rangle \qquad [1.1]$$

Here $S$ denotes the subsystem with finite degrees of freedom where the correlation and bonding effects, that is, the local or onsite screened full many body interaction, will be fully explored, and $E$ denotes the environment in which the sub-system is embedded. The Hamiltonians of $S$, $E$ and the interaction between the system and the environment are described below.

The full system Hamiltonian is

$$\widehat{\mathcal{H}} = \widehat{H} - \mu\widehat{N} = \widehat{\mathcal{H}}_S + \widehat{\mathcal{H}}_E + \sum_k t_k \widehat{\psi}_{S,k} \widehat{\psi}_{E,k}^\dagger + h.c. \qquad [1.2]$$

Here we have assumed that long range interactions such as Coulomb interactions between the electrons are captured by the $\widehat{\mathcal{H}}_S$ and $\widehat{\mathcal{H}}_E$. The coupling between $S$ and its environment $E$ is captured by the kinetic hopping term $\sum_k t_k \widehat{\psi}_{S,k} \widehat{\psi}_{E,k}^\dagger + h.c.$

This is a reasonable assumption because 1) the long-range classical Coulomb interaction is usually taken into account in a mean field manner and captured by the electrostatic field (a single particle term); 2) in the tight binding formulation of the strongly correlated system, it is common to consider only onsite interaction, and 3) it is possible to generalize the coupling so that operators in the coupling term are not just single fermion operators. The effective Hamiltonian can still be derived but the self-consistency condition is more complicated there.

We further assume that $\widehat{\mathcal{H}}_S$ and $\widehat{\mathcal{H}}_E$ contain only terms with an even number of fermion operators, and they each commute with the respective total number operators. This implies that we have the following commutation relationships

$$[\widehat{\mathcal{H}}_S, \widehat{\mathcal{H}}_E] = 0 \qquad [1.3]$$
$$[\widehat{\mathcal{H}}_S, \widehat{\psi}_{E,k}^\dagger] = 0, [\widehat{\mathcal{H}}_S, \widehat{\psi}_{E,k}] = 0 \qquad [1.4]$$
$$[\widehat{\mathcal{H}}_E, \widehat{\psi}_{S,k}^\dagger] = 0, [\widehat{\mathcal{H}}_E, \widehat{\psi}_{S,k}] = 0 \qquad [1.5]$$

and

$$\langle\Psi|\hat{\mathcal{H}}|\Psi\rangle = \langle\Psi_S|\hat{\mathcal{H}}_S|\Psi_S\rangle + \langle\Psi_E|\hat{\mathcal{H}}_E|\Psi_E\rangle + \sum_k t_k \langle\Psi_E|\langle\Psi_S|\hat{\psi}_{S,k}\hat{\psi}^\dagger_{E,k}|\Psi_S\rangle|\Psi_E\rangle + h.c. \quad [1.6]$$

Note that because of the anti-commutation relations of the fermion operators in *E* and *S*, we have in general

$$\hat{\psi}^\dagger_{E,k}|\Psi_S\rangle|\Psi_E\rangle = \hat{\zeta}|\Psi_S\rangle\hat{\psi}^\dagger_{E,k}|\Psi_E\rangle, \qquad \langle S,n_i|\hat{\zeta}|S,n_j\rangle = (-1)^{n_j}\delta_{i,j} \quad [1.7]$$

Here $n_i$ denotes the number of fermions in the state, and the operator $\hat{\zeta}_k$ is a diagonal matrix if the basis-functions for the local Hilbert space are eigenstates of the number operator $\widehat{N}_S$.

Solving the variational wavefunction $|\Psi_S\rangle$ amounts to finding the ground state of the effective Hamiltonian $\mathcal{H}^{Seff}_{i,j}$ for the subsystem *S*, with matrix elements given by

$$\mathcal{H}^{Seff}_{i,j} = \langle\psi^S_i|\hat{\mathcal{H}}_S|\psi^S_j\rangle + \sum_k t_k(-1)^{n_j}\langle\psi^S_i|\hat{\psi}_{S,k}|\psi^S_j\rangle \langle\Psi_E|\hat{\psi}^\dagger_{E,k}|\Psi_E\rangle$$
$$+ \sum_k t_k(-1)^{n_i} \langle\Psi_E|\hat{\psi}_{E,k}|\Psi_E\rangle\langle\psi^S_i|\hat{\psi}^\dagger_{S,k}|\psi^S_j\rangle \quad [1.8]$$

Note that we have traced away the degree of freedom associated with the environment *E*, and the effect of the environment results in the non-zero expectation values of $\hat{\psi}_{E,k}$ and $\hat{\psi}^\dagger_{E,k}$. We want to reiterate that this is the key difference, in our opinion, between our theory and DMET. As explained above, we have used the fact that $|\psi^S_i\rangle$ are eigenstates of $\widehat{N}_S$, and we have $\hat{\zeta} = (-1)^{n_j}$.

To close the self-consistency loop, we will link $\langle\Psi_E|\hat{\psi}^\dagger_{E,k}|\Psi_E\rangle$ with the corresponding subsystem average values $\langle\Psi_S|\hat{\psi}^\dagger_{S,k}|\Psi_S\rangle$, which usually are given by the symmetry of the total system that the separable form of the variational wavefunction is going to preserve, that is,

$$\langle\Psi_E|\hat{\psi}^\dagger_{E,k}|\Psi_E\rangle = f_k(\{\langle\Psi_S|\hat{\psi}^S_{k'}|\Psi_S\rangle\}) \quad [1.9]$$

and the simplest translational symmetry of a lattice will give

$$\langle\Psi_E|\hat{\psi}^\dagger_{E,k}|\Psi_E\rangle = \langle\Psi_S|\hat{\psi}^\dagger_{S,k'}|\Psi_S\rangle, \quad [1.9']$$

where $k'$ is related to $k$ by lattice translation. Note that without translational symmetry, we usually cannot assume that [1.9] has the form of [1.9']. For example, for an impurity problem, the exact calculation of the $\langle\Psi_E|\hat{\psi}^\dagger_{E,k}|\Psi_E\rangle$ will require some additional assumptions on the properties of the environment, which in some cases, is assumed to be a non-interacting

fermion reservoir. In that situation, the coherent entanglement is only local at the impurity site. Other examples may include bio-molecules embedded in water environment.

## 2-d Hubbard Model

We will use the 2-d square lattice Hubbard model as an example and use the nearest neighbor dimer as the subsystem, the minimal system that goes beyond single-site correlation.

For dimers, the local basis states are

$$
\begin{aligned}
|s,0\rangle &= |0\rangle|0\rangle, \\
|s,1\rangle &= |u\rangle|0\rangle = \psi_{1u}^{s\,\dagger}|s,0\rangle, \\
|s,2\rangle &= |0\rangle|u\rangle = \psi_{2u}^{s\,\dagger}|s,0\rangle, \\
|s,3\rangle &= |d\rangle|0\rangle = \psi_{1d}^{s\,\dagger}|s,0\rangle, \\
|s,4\rangle &= |0\rangle|d\rangle = \psi_{2d}^{s\,\dagger}|s,0\rangle, \\
|s,5\rangle &= |u\rangle|d\rangle = \psi_{1u}^{s\,\dagger}\psi_{2d}^{s\,\dagger}|s,0\rangle, \\
|s,6\rangle &= |d\rangle|u\rangle = \psi_{1d}^{s\,\dagger}\psi_{2u}^{s\,\dagger}|s,0\rangle, \\
|s,7\rangle &= |u\rangle|u\rangle = \psi_{1u}^{s\,\dagger}\psi_{2u}^{s\,\dagger}|s,0\rangle, \\
|s,8\rangle &= |d\rangle|d\rangle = \psi_{1d}^{s\,\dagger}\psi_{2d}^{s\,\dagger}|s,0\rangle, \\
|s,9\rangle &= |ud\rangle|0\rangle = \psi_{1u}^{s\,\dagger}\psi_{1d}^{s\,\dagger}|s,0\rangle, \\
|s,10\rangle &= |0\rangle|ud\rangle = \psi_{2u}^{s\,\dagger}\psi_{2d}^{s\,\dagger}|s,0\rangle, \\
|s,11\rangle &= |ud\rangle|u\rangle = \psi_{1u}^{s\,\dagger}\psi_{1d}^{s\,\dagger}\psi_{2u}^{s\,\dagger}|s,0\rangle, \\
|s,12\rangle &= |u\rangle|ud\rangle = \psi_{1u}^{s\,\dagger}\psi_{2u}^{s\,\dagger}\psi_{2d}^{s\,\dagger}|s,0\rangle, \\
|s,13\rangle &= |ud\rangle|d\rangle = \psi_{1u}^{s\,\dagger}\psi_{1d}^{s\,\dagger}\psi_{2d}^{s\,\dagger}|s,0\rangle, \\
|s,14\rangle &= |d\rangle|ud\rangle = \psi_{1d}^{s\,\dagger}\psi_{2u}^{s\,\dagger}\psi_{2d}^{s\,\dagger}|s,0\rangle, \\
|s,15\rangle &= |ud\rangle|ud\rangle = \psi_{1u}^{s\,\dagger}\psi_{1d}^{s\,\dagger}\psi_{2u}^{s\,\dagger}\psi_{2d}^{s\,\dagger}|s,0\rangle,
\end{aligned}
\qquad [2.1]
$$

Thus the self-consistent dimer effective Hamiltonian for a 2-d square lattice Hubbard model is

$$
\mathcal{H}_{i,j}^{Seff} = \left\langle s,i \left| \sum_{k=1,2}(U\hat{n}_{ku}\hat{n}_{kd} - \mu\hat{n}_k) + \sum_{\sigma=u,d}(t\hat{\psi}_{1\sigma}\hat{\psi}_{2,\sigma}^{\dagger} + t\hat{\psi}_{2\sigma}\hat{\psi}_{1,\sigma}^{\dagger}) \right| s,j \right\rangle \\
+ \sum_{\sigma=u,d} 3t(-1)^{n_j}\left(\langle s,i|\hat{\psi}_{1\sigma}|s,j\rangle\langle\hat{\psi}_{2\sigma}^{\dagger}\rangle + \langle s,i|\hat{\psi}_{2\sigma}|s,j\rangle\langle\hat{\psi}_{1\sigma}^{\dagger}\rangle\right) + h.c.
\qquad [2.2]
$$

Here, we have applied translational symmetry to arrive at the self-consistent condition (c.f. [1.9']). The periodic symmetry implies that the single dimer is *not* an impurity. The system $S$ maintains the translational symmetry of the lattice. There is a gauge degree of freedom related to basis wavefunctions that can have any phase. Consequently, $\langle \Psi^E | \hat{\psi}_{k\sigma}^\dagger | \Psi^E \rangle = \langle G_S^{eff} | \hat{\psi}_{k'\sigma}^\dagger | G_S^{eff} \rangle = \langle \hat{\psi}_{k'\sigma}^\dagger \rangle$, where $|G_S^{eff}\rangle$ is the ground state of $\mathcal{H}_{i,j}^{Seff}$ and the factor 3 in Eq. [2.2] is from 3 nearest neighbors.

We have found that an exact solution to the effective Hamiltonian has the following form and in a wide range of parameter space, it is the effective ground state:

$$|G_S^{eff}\rangle = \alpha\{|s,1\rangle - |s,2\rangle + |s,3\rangle - |s,4\rangle\} + \beta\{|s,5\rangle - |s,6\rangle\} + \gamma\{|s,9\rangle + |s,10\rangle\} \\ + \lambda\{|s,11\rangle - |s,12\rangle - |s,13\rangle + |s,14\rangle\} \quad [2.3]$$

All other terms are zero.

Note that the traditional Cooper pair off-diagonal order parameter is zero for both *s*-wave and *d*-wave channels:

$$\langle G_S^{eff} | \hat{\psi}_{1u}\hat{\psi}_{2d} | G_S^{eff} \rangle = \langle G_S^{eff} | \hat{\psi}_{2u}\hat{\psi}_{1d} | G_S^{eff} \rangle = \\ \langle G_S^{eff} | \hat{\psi}_{1u}\hat{\psi}_{1d} | G_S^{eff} \rangle = \langle G_S^{eff} | \hat{\psi}_{2u}\hat{\psi}_{2d} | G_S^{eff} \rangle = 0 \quad [2.4]$$

Instead, the off-diagonal non-zero order parameter is given by the non-zero expectation values of the following operator

$$\{(\hat{\psi}_{1u} - \hat{\psi}_{1d})\hat{\psi}_{2u}\hat{\psi}_{2d} - (\hat{\psi}_{2u} - \hat{\psi}_{2d})\hat{\psi}_{1u}\hat{\psi}_{1d}\}(\hat{\psi}_{1u}^\dagger - \hat{\psi}_{2u}^\dagger + \hat{\psi}_{1d}^\dagger - \hat{\psi}_{2d}^\dagger) \quad [2.5]$$

Note that this quadruple-fermion operator cannot be reduced to the conventional BCS order parameter operator form, which is a quadratic operator, or Cooper Pair operators, because of [2.4]. The authors also note that the *t-J* model (Zhang & Rice, 1988) physics is not the same as discovered in this paper. In *t-J* model, the superconductivity is still realized through coherent condensate of Cooper pairs of electrons/holes.

However, this surprise pales when compared to the realization that the self-consistent ground state solution contains the following single fermion off-diagonal "condensate":

$$\langle G_S^{eff} | \hat{\psi}_{1u} | G_S^{eff} \rangle = -\langle G_S^{eff} | \hat{\psi}_{2u} | G_S^{eff} \rangle = -\langle G_S^{eff} | \hat{\psi}_{1d} | G_S^{eff} \rangle = \langle G_S^{eff} | \hat{\psi}_{2d} | G_S^{eff} \rangle = \xi \quad [2.6]$$
$$\xi = -\alpha\beta + \alpha\gamma - \beta\lambda - \gamma\lambda$$

Also it is noted that for the exact half-filling case, $\alpha = \lambda = 0$, and we arrive at the quantum anti-ferromagnetic Neel State

$$|G_S^{AFM}\rangle = \beta\{|s,5\rangle - |s,6\rangle\} + \gamma\{|s,9\rangle + |s,10\rangle\} \quad [2.7]$$

The full exposition of these quantum phase transitions will be presented elsewhere.

## Finite Temperature Theory

The above self-consistent theory can be readily generalized to finite temperature with the local effective ground state replaced by the local density matrix and the self-consistency condition given by the expectation value with respect to the density matrix:

$$\hat{\rho}_S = \sum_i p_i |\Psi_i^{Seff}\rangle\langle\Psi_i^{Seff}|, \qquad [3.1]$$

$$\sum_j \mathcal{H}_{i,j}^{Seff} |\Psi_j^{Seff}\rangle = \varepsilon_i^S |\Psi_i^{Seff}\rangle \qquad [3.2]$$

$$p_i = \frac{\exp(\beta\varepsilon_i)}{\sum_k \exp(\beta\varepsilon_k)} \qquad [3.3]$$

## Conclusion

The potential application of the theory to the first principles quantum mechanical calculation of real materials, including large biomolecules such as DNA, is feasible due to the finite number of local degrees of freedom treated exactly by the proposed theory, as well as the mean field treatment of the interactions between the local system and its environment through the self-consistency condition. As demonstrated by this paper, the theory is able to shed light on some of the long-standing problems in many body physics, e.g., whether the Hubbard model provides the simplest model for superconductivity in doped narrow band oxide materials. A highly unconventional quadruple-fermion order parameter is discovered.

Acknowledgment: H.-P. Cheng acknowledges US DOE grant No. DE-FG02-02ER45995.

## References

Anderson, P. W. (1987). *Science 235*, 1196.

G.Kotliar, & Vollhardt, D. (2004). *Physics Today 57 (3)*, 53.

Ho, K., Schmalian, J., & Wang, C. (2008). *Phys. Rev. B 77 (7)*, 073101.

Knizia, G., & Chan, G. K.-L. (2012). Density Matrix Embedding: A Simple Alternative to Dynamic Mean Field Theory. *Phys. Rev. Lett. (109)*, 186404.

Knizia, G., & Chan, G. K.-L. (2013). Density Matrix Embedding: A Strong Coupling Quantum Embedding Theory. *J. Chem. Theory Comput. (9)*, 1428-1432.

Kohn, W., & Sham, L. J. (1965). *Phys. Rev. 140 (4A)*, 1133-1138.

Kotliar, G., Savrasov, S., Pallson, G., & Biroli, G. (2001). *Phys. Rev. Lett. 87*, 186401.

Sun, Q., & Chan, G. K.-L. (2016 ). Quantum Embedding Theories. *Accounts of Chemical Research (49)*, 2705-2712.

Zhang, F., & Rice, T. M. (1988). *Phys. Rev. B 37*, 3759.